\documentclass[12pt]{article}
\usepackage{graphicx}
\usepackage{amssymb}
\begin{document}
\begin{center}
{\Large\bf  The coincidence problem in $f(R)$ gravity models}\\[20mm]
A. Sil \footnote{St.Paul's C. M. College, 33/1 Raja Rammohan Sarani, Kolkata 700 009, India;\\
\indent email: amitavadrsil@rediffmail.com},   S. Som
\footnote{Meghnad Saha Institute of Technology, Nazirabad, East
Kolkata Township,\\ Kolkata 700 107,India;\indent email:
sumitsom79@yahoo.com}
\\[5mm]
{\em Relativity and Cosmology Research Centre,\\Department of Physics, Jadavpur University,\\
Kolkata - 700 032, India.} \\[20mm]
\end{center}

\pagestyle{myheadings}
\newcommand{\be}{\begin{equation}}
\newcommand{\ee}{\end{equation}}
\newcommand{\bea}{\begin{eqnarray}}
\newcommand{\eea}{\end{eqnarray}}

\begin{abstract}
To explore possibilities of avoiding coincidence problem in $f(R)$
gravity we consider models in Einstein conformal frame which are
equivalent to Einstein gravity with a minimally coupled scalar
field. As the conformal factor determines the coupling term and
hence the interaction between matter and dark energy, the function
$f(R)$ can in principle be determined by choosing an appropriate
function for the deceleration parameter only. Possible behavior of
$f(R)$ to avoid coincidence problem are investigated in
two such cases.
\end{abstract}

\vspace{0.5cm} Key Words: Cosmology; Dark energy; Coincidence Problem; $f(R)$ gravity.
\vspace{0.5cm}

\section{Introduction}
It is now more than a decade when the first observational evidence
indicated that our Universe might have been expanding with
acceleration\cite{riess}. With a growing number of data favoring
such evidences over past few years such doubt has now turned into
certainty that the Universe has entered an accelerated phase of
expansion from a decelerated phase very
recently\cite{cmbr}\cite{bao}. This observational fact however
needs a solid theoretical support which appears to be a major
challenge in cosmology today. Efforts to resolve this matter have
been made so far in two possible directions. One approach is to
introduce a new term in the energy-momentum side of the Einstein's
equation which will act as a source of gravity with repulsive
effect and initiate acceleration in the course of expansion. Such
a source has been termed in literature as `Dark Energy' due to its
unknown nature. It is usually parameterized by an equation of
state of the form $\omega_{DE} = p/\rho$. From Friedmann equation
one can see that a value of $\omega_{DE} < -1/3$ is required for
accelerated cosmic expansion. The most natural and simplest choice
as a candidate for dark energy could have been the Cosmological
Constant $\Lambda$ with equation of state $\omega _{DE}=-1$. But
if one believes that vacuum energy is the origin of it then one
fails to find any mechanism to obtain a value of $\Lambda$ that is
120 orders of magnitude less than the theoretical prediction to be
consistent with observation. Attempts were made by introducing the
concept of varying $\Lambda$ to explain observation but with same
equation of state $\omega _{DE}=-1$\cite{varying lambda}.
Possibilities of evolving $\omega _{DE}$ were also explored in
many dynamical dark energy models. Primary candidates in this
category are scalar field models such as
Quintessence\cite{quintessence} and K-essence\cite{k essence}. A
common example to quintessence is the energy of a slowly evolving
scalar field that has not yet reached the minimum of its potential
$V (\phi)$, similar to the inflaton field used to describe the
inflationary phase of the Universe. In quintessence models the
parameter range is $-1 < \omega_{DE} < -1/3$, and the dark energy
density decreases with a scale factor $a(t)$ as $\rho_{DE}\propto
a^{-3(1+\omega_{DE})}$\cite{turner}. A specific exotic form of
dark energy denoted phantom energy, with $\omega_{DE} < -1$, has
also been proposed\cite{Caldwell}. It possesses peculiar
properties, such as the violation of the energy conditions and an
infinitely increasing energy density. However, recent fits to
supernovae, cosmic microwave background radiation (CMBR) and weak
gravitational lensing data indicate that an evolving equation of
state crossing the phantom divide, is mildly favored, and several
models have been proposed in the literature\cite{phantom}. In
particular, models considering a redshift dependent equation of
state, possibly provide better fits to the most recent and
reliable SN Ia supernovae Gold dataset. A few comprehensive and
recent review articles on dark energy may be of interest in this
regard\cite{darkenergy}.

The alternative way to explain cosmic acceleration is to
anticipate that large scale dynamics of the Universe is not
governed by Einstein's equations. In a classical generalization of
general relativity, the Ricci scalar $R$ in Einstein-Hilbert
action is replaced by a more general function $f(R)$ which
modifies the geometry-side of the Einstein's equation\cite{fR}.
Earlier interest in $f(R)$ theories was motivated by inflationary
scenarios as for instance, in the Starobinsky model\cite{
starobinsky}. Recently, a number of models in $f(R)$ modified
theories of gravity has been verified in an attempt to explain the
late-time accelerated expansion of the Universe. In particular, it
has been shown that cosmic acceleration can indeed be explained in
the context of $f(R)$ gravity\cite{carrol}, and the conditions of
viable cosmological models have also been derived\cite{amendola}.
In the context of local tests in the Solar System regime, severe
weak field constraints however, seem to rule out most of the
models proposed so far\cite{solartest}, although viable models do
exist\cite{viablefR}. Some recent reviews on $f(R)$ gravity may
be of interest in this regard\cite{review_fR}.

\par There exist many cosmological models in `modified gravity' as
well as `dark energy models' that can successfully explain the
observed late-time acceleration. However, there is no strong
reason to choose one as better than the other. Not only that,
there remain many unsolved issues related to all these approaches.
One important issue among those is the Coincidence
Problem\cite{cp}. If we accept that the observed acceleration is
being caused by some `dark energy' it is striking to find out that
in our present Universe the density of dark energy and dark matter
are of the same order. In fact any cosmological model successful
in explaining late-time acceleration will necessarily suggest that
dark energy density has started dominating over matter energy very
recently. Now the question is why is it happening now? Is it a
mere coincidence or is there some deep underlying reason behind
it? This Coincidence Problem has been addressed in literature
extensively\cite{coinci}. One way to get rid of this is to design
a model in which the ratio of matter density to dark energy
density settles down with evolution to its present day value
quickly and remains stationary for a long period until today so
that the coincidence problem does not arise. However such a model
needs interaction between dark energy and matter. Attempts to
realize the coincidence problem as a consequence of interaction
between matter sector and dark energy are not new\cite{cps}. The
suggested interaction terms in all such attempts are however
phenomenological and are not derived from any fundamental theory.
In this present note we address the coincidence problem in the
context of $f(R)$ gravity models in Einstein conformal frame. The
conformally rescaled metric in such a model has a scalar field
partner that act as another dynamical variable of the vacuum
sector. Understandably this scalar field is not a kind of matter
field and is actually given in terms of $f(R)$. There exists a
coupling between this scalar degree of freedom and the matter
sector induced by this conformal transformation and the
interaction between matter and dark energy opens up the
possibility of addressing the coincidence problem.   The field
equations look very much like those of quintessence models but
with a difference. The interaction term is given by the conformal
factor and hence the coupling constant is universal in a sense
that the scalar field couples with same strength to all kinds of
matter fields. In a recent work Bisabr\cite{bisabr} claims to have
derived necessary conditions for alleviating coincidence problem
in some $f(R)$ gravity models in Einstein frame. In these models
the functional forms of $f(R)$ are chosen by hand and the
parameters in it are fixed by putting constraints to make the
models free from coincidence problem and produce the present day
observed value of deceleration parameter. We noticed however that
the choice of $f(R)$ fixes the dynamics of the model and although
it produces desired value of present day deceleration parameter
correctly; its time evolution appears to be not in agreement with
observation. We suggest in this paper a different approach to
overcome this difficulty. We take up two models by choosing the
functional forms of deceleration parameter from those available in
literature and look for a suitable $f(R)$ that can alleviate the
coincidence problem. In the first model the possible asymptotic
behavior of $f(R)$ that can solve the problem is shown. The second
model however clearly rules out the existence of any well behaved
$f(R)$ for this purpose.

\section{Action and field equations}
In the $f(R)$ modified theories of gravity, the standard
Einstein-Hilbert action is replaced by an arbitrary function of
Ricci scalar $R$ and is given by
\begin{equation}\label{actionfR}
 S=\frac{1}{2k}\int d^{4}x \sqrt{-g}f(R)+S_{M}(g^{\mu\nu},\psi),
\end{equation}
where $k=8\pi G$. The matter action $S_{M}(g^{\mu\nu},\psi)$ is
defined as \[S_M=\int d^{4}x \sqrt{-g}
\mathcal{L}_m(g^{\mu\nu},\psi),\] where $\mathcal{L}_m$ is the
matter Lagrangian density and $\psi$ denotes all matter fields
collectively. Using metric formalism, by varying the above action
with respect to $g^{\mu\nu}$, one can arrive at the field
equations

\begin{equation}\label{fe}
 f' R_{\mu\nu}-\frac{1}{2}fg_{\mu\nu}-\nabla_{\mu}\nabla_{\nu}f'+g_{\mu\nu}\Box f'=k T^{(m)}_{\mu\nu},
\end{equation}
where $f'\equiv df/dR$. The matter stress-energy tensor
$T^{(m)}_{\mu\nu}$, is defined as
\begin{equation}\label{tmunu}
    T^{(m)}_{\mu\nu}=-\frac{2}{\sqrt{-g}}\frac{\delta (\sqrt{-g}\mathcal{L}_{m})}{\delta (g^{\mu\nu})}.
\end{equation}
The $f(R)$ gravity may be written as a scalar-tensor theory, by
introducing a Legendre transformation $\{R,f\}\rightarrow
\{\phi,U\}$ defined as
\begin{equation}\label{lt}
    \phi\equiv f'(R),   U(\phi)\equiv R(\phi)f'-f(R(\phi)).
\end{equation}
In this representation the field equations of $f(R)$ gravity can
be derived from a Brans-Dicke type action given by
\begin{equation}\label{bd}
 S=\frac{1}{2k}\int d^{4}x \sqrt{-g}[\phi R-U(\phi)+\mathcal{L}_m].
\end{equation}
This is the Jordan frame representation of the action of a
Brans-Dicke theory with the parameter $\omega =0$. As usual in
Brans-Dicke theory and more general scalar-tensor theories, one
can perform a canonical transformation and rewrite the action
(\ref{bd})in what is called Einstein frame. Rescaling the metric
as
\begin{equation}\label{ct1}
  g_{\mu\nu}\rightarrow\tilde{g}_{\mu\nu}=f'(R)g_{\mu\nu}
\end{equation}
and redefining $\phi \rightarrow \tilde{\phi}$ with
\begin{equation}\label{ct2}
    d\tilde{\phi}=\sqrt{\frac{3}{2k}}\frac{d\phi}{\phi}
\end{equation}
the scalar-tensor theory can be mapped into the Einstein frame in
which the `new' scalar field $\tilde{\phi}$ couples minimally to
the Ricci curvature and has canonical kinetic energy as described
by the gravitational action
\begin{equation}\label{action}
  S=\int d^{4}x \sqrt{-g}\left[ \frac{\tilde{R}}{2k}-\frac{1}{2}\partial^{\alpha}\tilde{\phi}\partial_{\alpha}\tilde{\phi}-V(\tilde{\phi})\right] +S_{M}(e^{-2\beta\tilde{\phi}}\tilde{g}^{\mu\nu},\psi).
\end{equation}
The self-interacting potential $V(\tilde{\phi})$ is given by
\begin{equation}\label{potential}
  V(\tilde{\phi})=\frac{Rf'(R)-f(R)}{2k f'^{2}(R)}.
\end{equation}
Apparently, a coupling of the scalar field $\tilde{\phi}$ with the
matter sector sector is now induced. The strength of this coupling
$\beta=\sqrt{\frac{1}{6}}$ is however, fixed and is same for all
matter fields. Taking $\tilde{g}_{\mu\nu}$ and $\tilde{\phi}$ as
two independent field variables, the variations of the action
(\ref{action})yield the following field equations respectively
\begin{eqnarray}
\tilde{G}_{\mu\nu} &=& k(\tilde{T}^{\tilde{\phi}}_{\mu\nu}+\tilde{T}^{m}_{\mu\nu}) \label{fe1}\\
\Box \tilde{\phi}-\frac{dV(\tilde{\phi})}{d\tilde{\phi}} &=&
-\beta\sqrt{k}\tilde{T}^{m}\label{fe2}
\end{eqnarray}
where
\begin{eqnarray}
  \tilde{T}^{\tilde{\phi}}_{\mu\nu} &=& \nabla_{\mu}\tilde{\phi}\nabla_{\nu}\tilde{\phi}- \frac{1}{2}\tilde{g}_{\mu\nu}\nabla^{\alpha}\tilde{\phi}\nabla_{\alpha} \tilde{\phi}-V(\tilde{\phi})\tilde{g}_{\mu\nu}\\
  \tilde{T}^{m}_{\mu\nu} &=& \frac{-2}{\sqrt{-g}}\frac{\delta S_{M}(\tilde{g}^{\mu\nu},\psi)}{\delta (\tilde g^{\mu\nu})}
\end{eqnarray}
are stress-tensors for the scalar field and the matter field. It
is important to note that $\tilde{T}^{\tilde{\phi}}_{\mu\nu}$ and
$\tilde{T}^{m}_{\mu\nu}$ are not separately conserved and hence
satisfy the following equation
\begin{equation}
   \tilde{\nabla}^{\mu}\tilde{T}^{m}_{\mu\nu}= -\tilde{\nabla}^{\mu}\tilde{T}^{\tilde{\phi}}_{\mu\nu}= \beta\sqrt{k}\nabla_{\nu}\tilde{T}^{m}.
\end{equation}
The trace of matter stress-tensor $\tilde{T}^{m}$ is obtained by
taking trace of the field equation (\ref{fe1})
\begin{equation}
    \tilde{T}^{m}=\nabla^{\alpha}\tilde{\phi}\nabla_{\alpha} \tilde{\phi}+ 4V(\tilde{\phi})-\tilde{R}/k.
\end{equation}
Hereafter in the rest of this article we shall stop using tilde
overhead for convenient reading.

For a spatially flat homogeneous isotropic Universe described by
the Robertson-Walker metric
\begin{equation}\label{metric}
    ds^2=-dt^2 +a^2(dr^2+r^2d\theta^2+r^2\sin^2\theta d\phi^2)
\end{equation}
assuming matter in the form of pressureless dust the field
equations (\ref{fe1}) reduce to
\begin{eqnarray}
  3H^2 &=& k(\rho_{m}+\rho_{\phi}) \label{feq1}\\
 \mbox{and}\hspace{1cm}2\dot{H}+3H^2 &=& -k\omega_{\phi}\rho_{\phi}.\label{feq2}
\end{eqnarray}
Whereas field equation (\ref{fe2}) reduces to
\begin{equation}\label{wave}
    \ddot{\phi}+3H\dot{\phi}+\frac{dV}{d\phi}=-\beta\sqrt{k}\rho_{m}.
\end{equation}
Here an over dot denotes derivative with respect to time $t$. The
character $\rho$ denotes the energy density of matter or other
fluids. Throughout this article suffix $m$ to a quantity denotes
its matter part while the suffix $\phi$ denotes the `dark energy'
part of it. $H$ is the usual Hubble parameter defined as $H \equiv
\dot{a}/a$ and $\omega_{\phi}= \frac{p_{\phi}}{\rho_{\phi}}$ is
the equation of state parameter of the scalar field $\phi$. The
density $\rho_{\phi}$ and pressure $p_{\phi}$ of the scalar field
are defined as
\begin{eqnarray}
  \rho_{\phi} &=& \frac{1}{2}\dot{\phi}^2+V(\phi) \nonumber\\
  \mbox{and}\hspace{1cm}p_{\phi} &=& \frac{1}{2}\dot{\phi}^2-V(\phi)\nonumber
\end{eqnarray}
\par Due to the coupling between the scalar field and matter field,
the two components interact and hence are not separately
conserved. From (\ref{feq1}), (\ref{feq2}) and (\ref{wave}) the
conservation equations for matter and the scalar field follow
easily to give
\begin{eqnarray}
    \dot{\rho_{m}}+3H\rho_{m}=Q \label{conservation1}\\
    \mbox{and}\hspace{1cm}\dot{\rho_{\phi}}+3H(1+\omega_{\phi})\rho_{\phi}=-Q\label{conservation2}
\end{eqnarray}
where the interaction term
\begin{equation}
Q=\beta\sqrt{k}\dot{\phi}\rho_{m}\label{Q}
\end{equation}

\par This implies that for
$\dot{\phi}>0$, $Q$ and hence $\dot{\rho_{m}}$ is positive and
energy flows from dark energy to dark matter and for
$\dot{\phi}<0$ energy is transferred in the opposite direction. It
is interesting to note that the interaction term (\ref{Q}) looks
similar to some of the phenomenological coupling terms suggested
in literature\cite{cps}. It vanishes when $\phi=$ constant
\emph{i.e.} when  $f(R)$ is linear. A vanishing $Q$ implies that
matter and dark energy remain separately conserved.

\section{Outline Of The Present Work}

To avoid the coincidence problem, matter and dark energy must
scale each  other over a considerably long period of time during
the later stage of evolution of the Universe. In other words, the
ratio of two energy densities
$r\equiv\frac{\rho_{m}}{\rho_{\phi}}$ remains constant in spite of
their different rates of time evolution. Thus an interaction
between the matter and dark energy becomes necessary in these kind
of models. An appropriate choice of the interaction term may lead
towards this goal but usually such a choice is phenomenological
and does not come from any fundamental theory.  However, in $f(R)$
gravity models in the Einstein frame, the form of $Q$ is
determined by the transformation itself and one does not have the
liberty of tailoring the interaction. Thus to achieve  a
stationary value for $r$ and hence to avoid the coincidence
problem, we can look for an appropriate form of $f(R)$ instead.

\par The set of three independent equations (\ref{feq1}), (\ref{feq2}) and (\ref{wave})
contains four unknowns viz: $H$, $\rho_{m}$, $\phi$ and $V(\phi)$. Choice of any one of
these parameters (or a derived quantity like the deceleration
parameter $q$) shall in principle suffice to determine the
functional form of all other quantities including $f(R)$ and $Q$
and one may check how well the model alleviates the coincidence
problem fitting adequately with observations. In contrast, one can
also start with a choice of $f(R)$ and solve for the other
parameters of the model using the set of equations presented in
the previous section and check with observations. However, due to
complexities involved in the field equations this approach is
difficult to work out. In the present work we take the former
approach i.e search for a function $f(R)$ for a given form of $q$
such that the ratio of energy densities $r$ settles at a
stationary value in course of evolution.

\par The Hubble parameter $H(t)\equiv\dot{a}/a$ and the deceleration
parameter $q(t)\equiv\frac{-\ddot{a}}{(aH)^2}$ are related as
functions of the redshift $z (=-1+\frac{a_{0}}{a})$ by the
equation,

\begin{equation}\label{Hz}
    H(z)= H_{0} \textsl{exp}\left[\int_{0}^{z}[1+q(z')]d\ln(1+z')\right]
\end{equation}
where the subscript $0$ refers to the present day values of the
variables. Also note that by definition,
\begin{equation}\label{Hdot}
    \dot{H}= -(1+q)H^2.
\end{equation}

\noindent Finally, the Ricci scalar $R$ can be expressed in terms
of $q$ and $H$ as

\begin{equation}\label{ricci}
    R = 6(1-q)H^2.
\end{equation}

On the other hand, let us consider the time evolution of the ratio
$r\equiv\frac{\rho_{m}}{\rho_{\phi}}$,

\begin{equation}\label{rdot1}
    \dot{r}=\frac{\dot{\rho_{m}}}{\rho_{\phi}}-r\frac{\dot{\rho_{\phi}}}{\rho_{\phi}}.
\end{equation}

\noindent Using equations (\ref{conservation1}),
(\ref{conservation2}) and (\ref{Q}) in (\ref{rdot1}) we obtain,

\begin{equation}\label{rdot2}
    \dot{r}= 3Hr\omega_{\phi}+ \beta\sqrt{k}\dot{\phi}r(1+r)
\end{equation}

\noindent Application of (\ref{Hdot}) in equation (\ref{feq2})
leads to

\begin{equation}\label{omega}
    \omega_{\phi}=\frac{2q-1}{k\rho_{\phi}}.
\end{equation}

\noindent Substituting (\ref{omega}) in (\ref{rdot2}) and using
equation (\ref{feq1}) we arrive at

\begin{equation}\label{rdot3}
\frac{\dot{r}}{r(1+r)}= (2q-1)H + \beta\sqrt{k}\dot{\phi}.
\end{equation}

We can show that the observed equality in the order of magnitude
of the dark energy density and matter density of the Universe in
recent epochs is not merely a coincidence if we demand that the
ratio of the densities $r$ either remain constant or vary
extremely slowly as compared to the scale factor $a$ over a long
period of time. This means to analyze recent behavior of the
Universe one can safely assume $\dot{r}=0$ and therefore from
(\ref{rdot3}) we can write

\begin{equation}\label{rdot4}
(2q-1)H = -\beta\sqrt{k}\dot{\phi}.
\end{equation}

\noindent Replacing $\dot{\phi}$ in (\ref{rdot4}) by using
equation (\ref{ct2}) gives

\begin{equation}\label{fr}
\frac{f''}{f'}\dot{R}=-2(2q-1)H
\end{equation}
where a prime denotes derivative with respect to $R$.\\

\par Thus for a given $q(z)$ one can easily determine the evolution
of the hubble parameter $H$ and the Ricci scalar $R$ as functions
of the redshift $z$, using equations (\ref{Hz}) and (\ref{ricci}).
These functions can then be used to solve equation (\ref{fr}) and
a functional form of $f(R)$, valid in the $\dot{r}=0$ regime,
can in principle be determined . In the subsequent sections we
make attempts to determine $f(R)$ for two different parametric
choices of $q(z)$ following the above prescription to alleviate
coincidence problem.

\section{Model I}\label{M1}

The simplest parametrization of the deceleration parameter $q$ in
terms of the redshift $z$ available in literature is the linear
two parameter relation given by\cite{reiss2004}\cite{cunha}

\begin{equation}\label{qz1}
q(z)=q_{0}+ q'z
\end{equation}

\noindent where $q_{0}=q|_{z=0}$ is the present value of the
deceleration parameter and $q'=\frac{dq}{dz}|_{z=0}$. For a
spatially flat Robertson-Walker Universe the best fit to the pair
of parameters is ($q_{0}$, $q'$) = (-0.73, 1.5) and the transition
redshift for these set of values is $z_{t}=0.49$\cite{cunha}. Note
that for $z \geq 1.154$, the deceleration parameter $q>1$. Such a
parametrization is therefore justified only at low redshifts over
a small range $0 \leq z \leq 1.15$. Assuming $q(z)$ as in (\ref{qz1}),
equations (\ref{Hz}) and (\ref{ricci}) give

\begin{eqnarray}
  H(z)= H_{0}\exp[q'z](1+z)^{(1+q_{0}-q')} \label{Hz1}\\
 \mbox{and}\hspace{1cm}R(z)=6H_{0}^{2}\exp[2q'z](1-q_{0}-q'z)(1+z)^{2(1+q_{0}-q')}\label{R1}
\end{eqnarray}

\noindent Integrating (\ref{fr}) and by using (\ref{qz1}) and
(\ref{Hz1}), we get

\begin{equation}
\frac{df}{dz}=f'_{0}\exp[4q'z](1+z)^{(4q_{0}-4q'-2)}\frac{dR}{dz}\label{fz1}
\end{equation}

\noindent where $f'_{0}=\frac{df}{dR}|_{z=0}$. Note that since the functional form of $q(z)$ in this model is valid over
a small range $0 \leq z \leq 1.15$, therefore the functional behavior of $f(R)$ as depicted in the above equation is only
asymptotic and may not indicate its true behavior for $z>>1$.
\par
\begin{figure}
\begin{center}
\includegraphics[width=6cm]{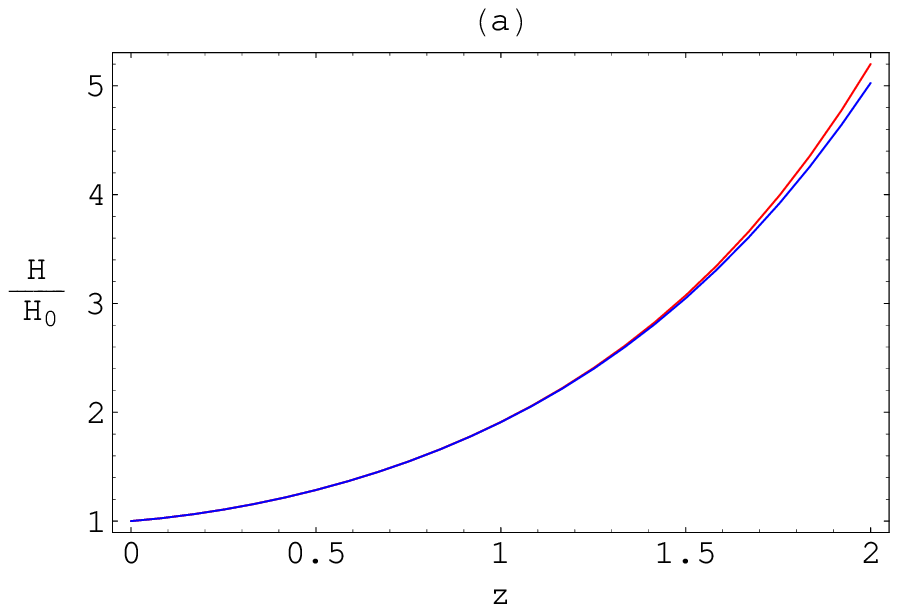}
\includegraphics[width=6cm]{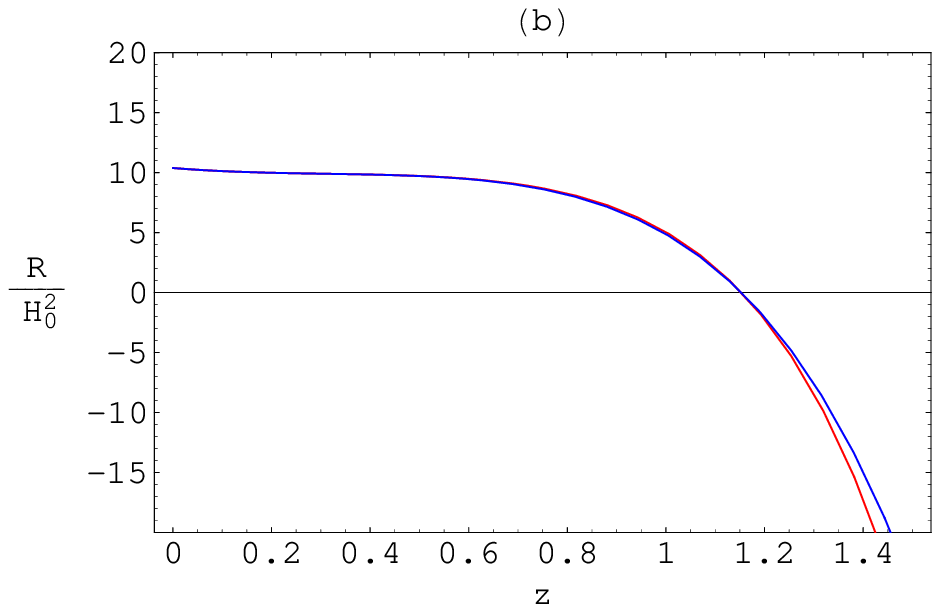}\\
\includegraphics[width=6cm]{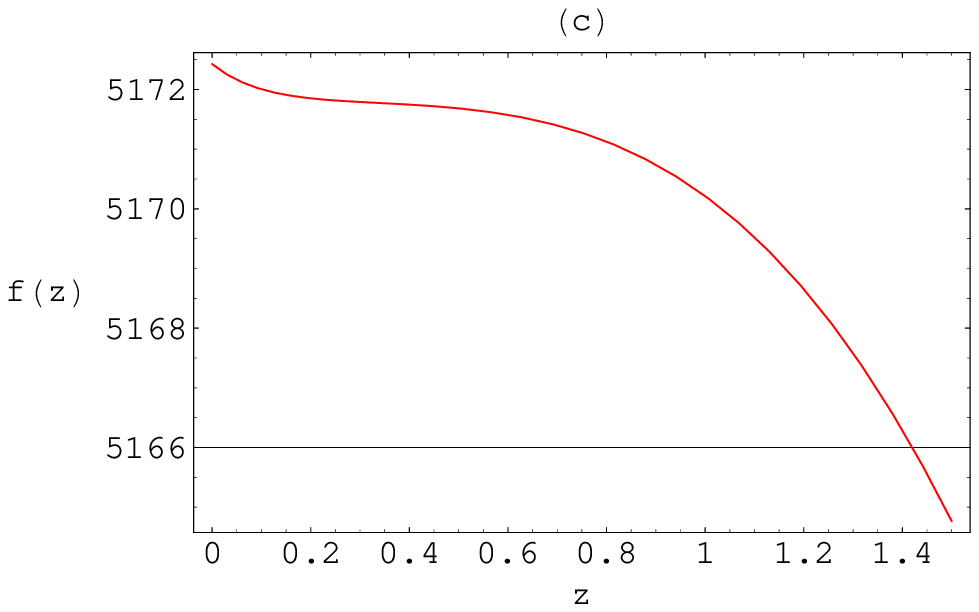}
\includegraphics[width=6cm]{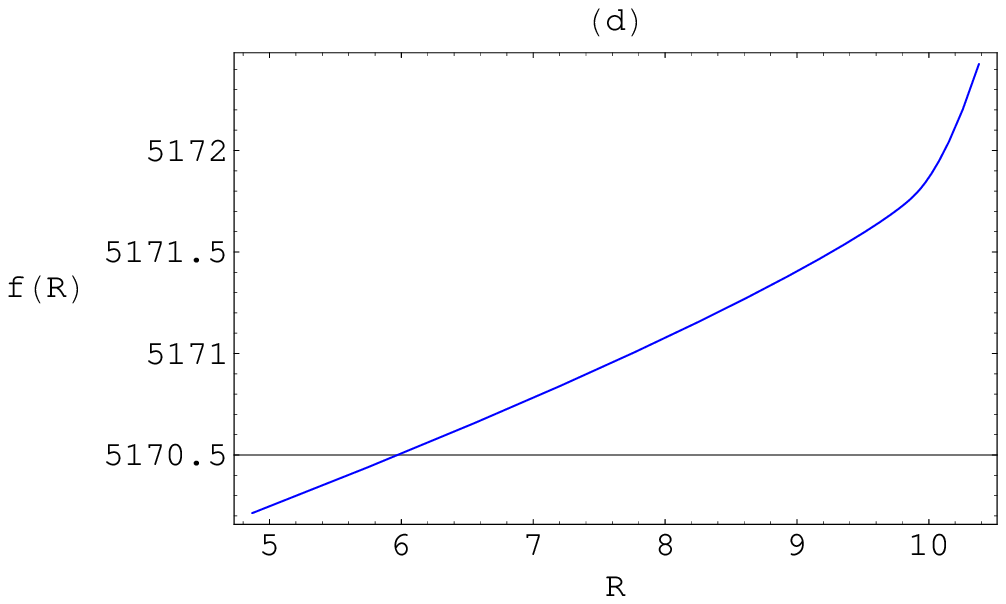}
\caption{\it The plots $(a)$ and $(b)$ show the variation of
actual and approximate expressions of $H(z)$ and $R(z)$ vs. $z$ in
red and blue respectively for model I with $q_{0}=-0.73$ and
$q'=1.5$. The curves appear to match exactly at low redshifts.
Variation of $f(z)$ vs. $z$ and $f(R)$ vs. $R$ are shown in $(c)$
and $(d)$ respectively with $f'_{0}=2$ and $H_{0}=1$
units.}\label{approx1}
\end{center}
\end{figure}

 The complexity involved in the expression on the r.h.s of
(\ref{fz1}) does not permit an analytical solution for $f(z)$.
However, since the expressions are true only for small $z$ we can always look for an approximation. Note that the expressions of $H(z)$, $R(z)$ and
$\frac{df}{dz}$ given by (\ref{Hz1}), (\ref{R1}) and (\ref{fz1})
contain an exponential term each. Retaining the terms up to the
sixth power of $z$ in the series expansion of the exponentials and
using them in (\ref{fz1}) an analytical form of $f(z)$ can be
obtained.  The analytical expression, however, is not being
mentioned here for economy of space. Instead variation of $f(z)$
versus $z$ under the above approximation is plotted in
fig.\ref{approx1}. That such an approximation holds good in our
range of interest ($0 \leq z \leq 1.15$) is also clear from
fig.\ref{approx1}(a)and (b) where we find that the curves for
$H(z)$ and $R(z)$ drawn with and without the approximation appears
almost indistinguishable at low redshifts. A parametric plot of
$R$ and $f$ with $z$ as parameter, is then plotted in
fig.\ref{approx1}(d)to obtain the nature of variation of $f(R)$
against $R$.
\par
\begin{figure}
\begin{center}
\includegraphics[width=6cm]{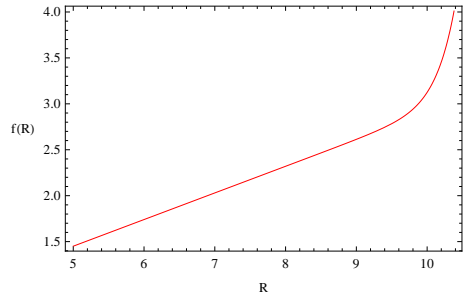}
\caption{\it The curve shows the plot of $f(R)$ vs.$R$ as given in
equation (\ref{trialf}) for $\lambda_1 = 0.29$, $\lambda_2 = 2.25
\times10^{-41}$ and $n = 40$.}\label{parplot1}
\end{center}
\end{figure}

 Thus in order to address the coincidence problem
in this model the function $f(R)$ should be such that, at low
redshift it behaves like the curve given in fig.\ref{approx1}(d).
The plot shows that the function $f(R)$ which is initially linear
in nature for $z\geq0.5$ implying no interaction with matter
sector, attains a non linear character in the very recent past
$z\leq0.5$ facilitating the interaction and hence a transition
from deceleration to an accelerated phase of expansion of the
Universe. It is always possible to suggest an analytic function
that can match the behavior of $f(R)$ as plotted in fig.1(d) but
we must remember that any such choice that fits the curve does not
guarantee $\dot{r}=0$. For example the function
\begin{equation}\label{trialf}
f(R)=\lambda_{1}R+\lambda_{2}R^{n}
\end{equation}
when plotted against $R$ appears to behave similarly (shown in fig.\ref{parplot1}).

The behavior of $\dot{r}$ for the above choice of $f(R)$ however deviates from desirable features. For this reason we refrain from suggesting any $f(R)$ but stress on the fact that such a function can always be found in principle that not only fits the curve but also guarantees $\dot{r}=0$, thereby alleviating the coincidence problem.

\section{Model II}

Another parametrization of $q(z)$ common in literature is of the
form\cite{gong1}

\begin{equation}\label{qz2}
q(z)=\frac{1}{2}+ \frac{q_{1}z + q_{2}}{(1+z)^2}
\end{equation}

\noindent where $q_{1}$ and $q_{2}$ are constants with
($\frac{1}{2}+q_{2}$) giving the present value of the deceleration
parameter and $\frac{dq}{dz}|_{z=0}=q_{1}- 2q_{2}$. For a
spatially flat Robertson-Walker Universe the best fit to the pair
of parameters is ($q_{1}$, $q_{2}$) = (1.47, -1.46) and the
transition redshift for these set of values is
$z_{t}=0.36$\cite{gong2}. For high redshifts $z\gg1$, the
deceleration parameter $q$ remains fixed at $\frac{1}{2}$ in this
model.

Proceeding in exactly the same manner as in section (\ref {M1}) we
get

\begin{eqnarray}
  H(z)=H_{0}(1+z)^{\frac{3}{2}}\exp\left[\frac{q_{2}}{2}+\frac{q_{1}z^{2}-q_{2}}{2(1+z)^{2}}\right] \label{Hz2}\\
 R(z)=6H_{0}^{2}(1+z)^{3}\left(\frac{1}{2}-\frac{q_{1}z+q_{2}}{(1+z)^{2}}\right)\exp\left[q_{2}+\frac{q_{1}z^{2}-q_{2}}{(1+z)^{2}}\right]\label{R2}\\
 \mbox{and}\hspace{1cm}\frac{df}{dz}=f'_{0}\exp\left[\frac{-2(2q_{1}z+q_{1}+q_{2})}{(1+z)^{2}}\right]\frac{dR}{dz}\label{fz2}
\end{eqnarray}

\begin{figure}
    \begin{center}
     \includegraphics[width=6cm]{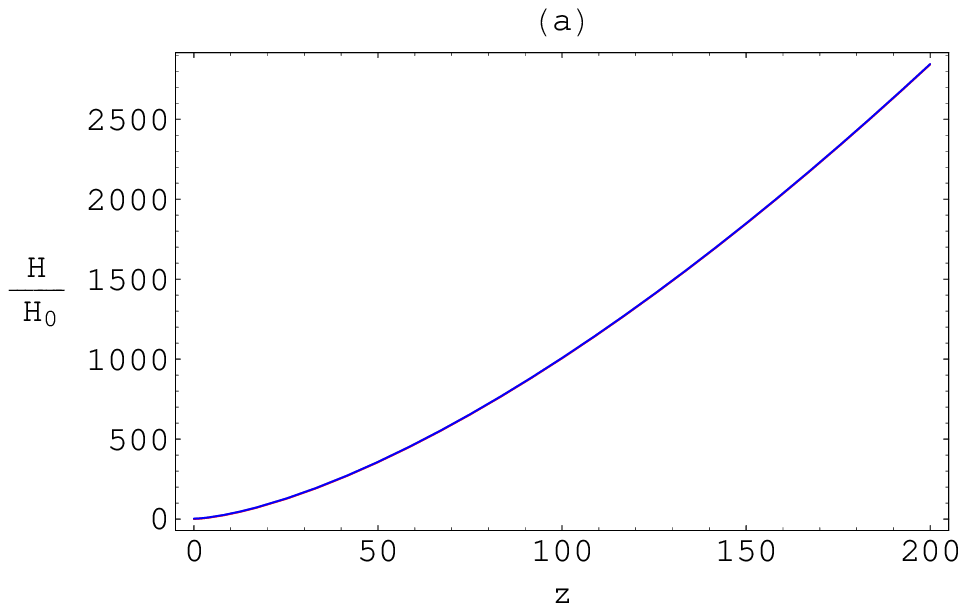}
     \includegraphics[width=6cm]{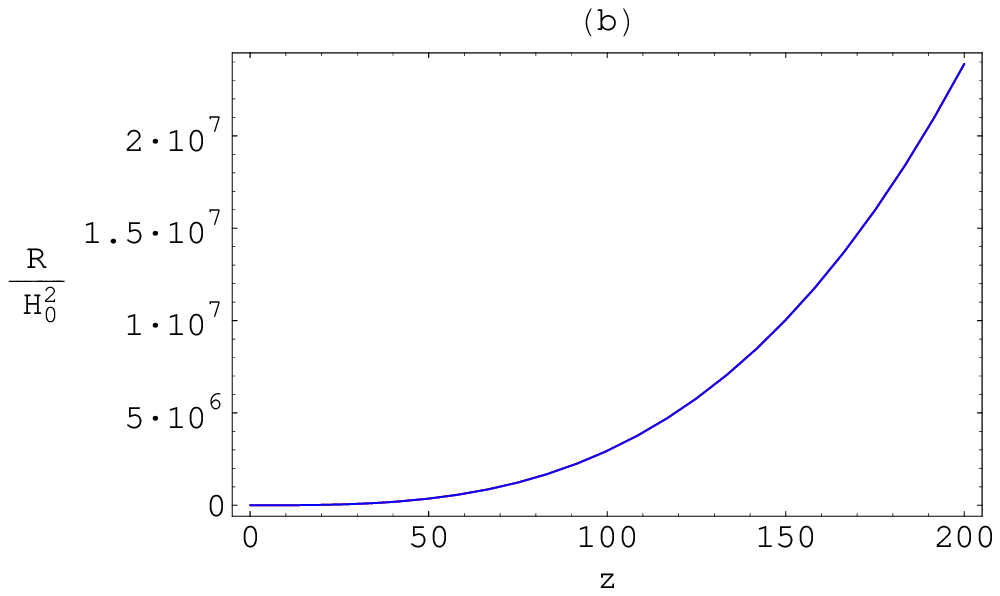}\\
     \includegraphics[width=6cm]{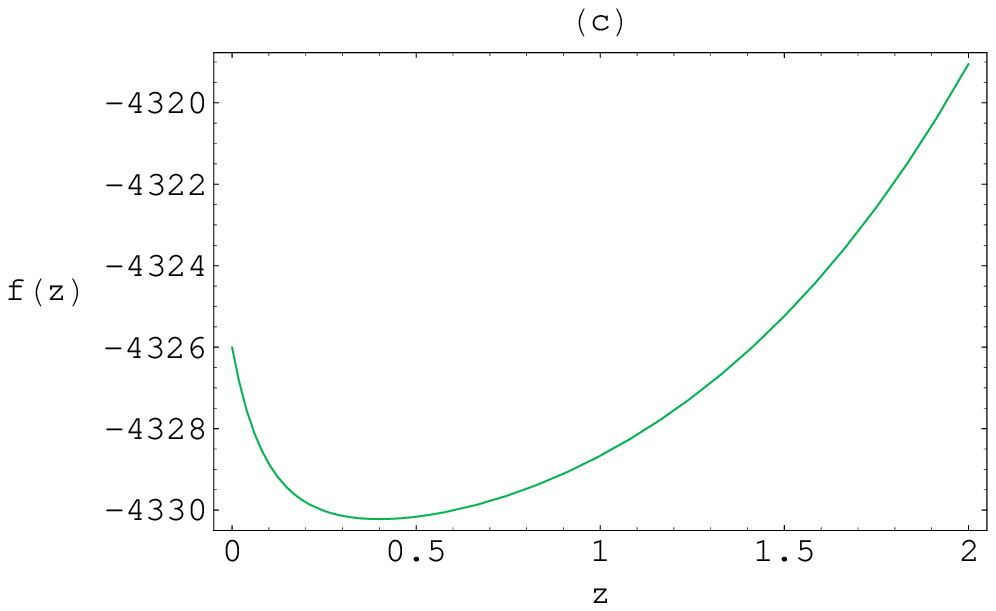}
  \caption{\it The plots $(a)$ and $(b)$ show the variation of actual and approximate expressions
  of $H(z)$ and $R(z)$ vs. $z$ in red and blue respectively for model II with $q_{1}=1.47$ and $q_{2}=-1.46$.
  The curves appear to match exactly over a long range of redshift. Variation of $f(z)$ vs. $z$ is shown in $(c)$ for
   $f'_{0}=2$ and $H_{0}=1$ units.}\label{approx2}
  \end{center}
\end{figure}

\begin{figure}
    \begin{center}
          \includegraphics[width=6cm]{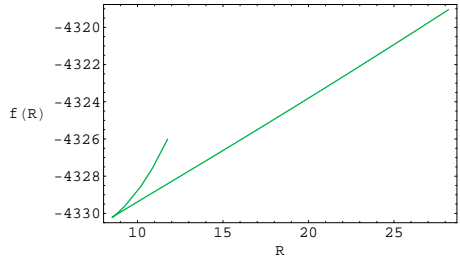}
  \caption{\it Plot of $f(R)$ vs. $R$ for model II with $q_{1}=1.47$, $q_{2}=-1.46$ and $f'_{0}=2$ in $H_{0}=1$ units.}\label{parplot2}
 \end{center}
\end{figure}

Obtaining an analytical solution for $f(z)$ from equation
(\ref{fz2}) is difficult in this case as well. However, following
similar approximation in the exponential terms of the above
equations, as in the previous model, a solution for $f(z)$ can be
easily obtained and the nature of the function
is shown in fig.\ref{approx2} along with the variations of
$H(z)$ and $R(z)$ against $z$. The plots of $H(z)$
and $R(z)$ with and without the approximations are found to be
almost indistinguishable over a long range of $z$ justifying the approximation. Nevertheless, the parametric plot in
fig.\ref{parplot2} shows that the function $f(R)$ in this case
is double valued in the recent phase of evolution of the Universe.

\par Therefore, for the particular parametrization of $q(z)$ chosen in
(\ref{qz2}), one cannot obtain a suitable well behaved function
$f(R)$, capable of alleviating the coincidence problem. This
result in contrast with that obtained in\cite{bisabr} can be
understood if we notice that a prior choice of $f(R)$ fixes the
functional behavior of  $q$ and hence that of $q'$. In\cite{bisabr} $f(R)$ is chosen by hand and the present value of
$q$ is used to fix the parameters in the given $f(R)$, but it
fails to achieve the correct value of $q'|_{z=0}$ in the process.
Our approach on the other hand, is just opposite. We fix the
function $q$ that fits with observational data and look for a
suitable $f(R)$ that can alleviate the coincidence problem. We
find that no such $f(R)$ exists as long we restrict ourselves to
$q$ that behaves as given in (\ref{qz2}).

\section{Conclusions}

A suitable kind of interaction between dark energy and dark matter
can make the ratio of their densities possible to attain a
stationary value during the course of evolution and consequently
can help alleviating the coincidence problem. In Einstein gravity
such interactions are phenomenological and needs fine tuning of
the interaction term. On the other hand, in Einstein frame
representation of $f(R)$ gravity models, the interaction between
the dark energy (scalar partner of the metric tensor) and dark
matter is fixed by the conformal transformation. In this present
article we address coincidence problem in the context of $f(R)$
gravity in Einstein frame. We point out that fixing the functional
form of any one of the four quantities viz. the Hubble parameter,
the matter density, the scalar field or the potential (or any
derived quantity like the deceleration parameter ) shall in
principle suffice to determine the functional form of all other
quantities including $f(R)$. We choose two functional forms for
deceleration parameter readily available in literature those fit
observational data. The linear function $q(z)$ in Model I
determines the the functional behavior of $f(R)$ for low redshift
which is plotted in fig.\ref{approx1}(d). The plot shows that the
function $f(R)$ which is initially linear in nature for $z\geq0.5$
implying no interaction with matter sector, attains a non linear
character in the very recent past $z\leq0.5$ facilitating the
interaction and hence a transition from deceleration to an
accelerated phase of expansion of the Universe. In another attempt
a popular non-linear function of $q(z)$ is chosen for
investigation. However the function $f(R)$ obtained in this case
is double valued at low redshifts and hence is not suitable from
physical point of view. Thus our investigation rules out the
possibility of such form of deceleration parameter in $f(R)$
gravity if one is looking for a model free from coincidence
problem.

\end{document}